\documentclass[twocolumn,showpacs,tighten,floats,amsmath,amssymb,prb]{revtex4}
\usepackage{graphicx}    
\usepackage{epsfig}      
\usepackage{dcolumn}     
\usepackage{bm}          
\newcommand{\apst}{AgPb$_{m}$SbTe$_{m+2}$}
\newcommand{\twelve}{AgPb$_{x}$SbTe$_{x+2}$}
\newcommand{\one}{AgSbTe$_{2}$}
\newcommand{\two}{AgPb$_{6}$SbTe$_{8}$}
\newcommand{\four}{AgPb$_{12}$SbTe$_{14}$}
\newcommand{\six}{Ag$_{0.86}$Pb$_{18}$SbTe$_{20}$}
\newcommand{\seven}{PbTe}

\newcommand{\nine}{AgPb$_{2}$SbTe$_{4}$}
\newcommand{\ten}{AgPbSbTe$_{3}$}
\newcommand{\eleven}{AgPb$_{3}$SbTe$_{5}$}
\newcommand{\thirteen}{AgPb$_{4}$SbTe$_{6}$}

\begin{document}

\title{Nanoscale clusters in the high performance thermoelectric \apst\ }
\author {H. Lin,  E.~S. Bo\v{z}in,  S.~J.~L. Billinge}
\affiliation{Department of Physics and Astronomy, Michigan State
University, East Lansing, MI 48824}
\author{Eric Quarez, M.~G. Kanatzidis}
\affiliation{Department of Chemistry, Michigan State University,
East Lansing, MI 48824}
\email{billinge@pa.msu.edu}
\homepage{http://nirt.pa.msu.edu/}
\date{\today}
\begin{abstract}
   The local structure of the \apst\ series of thermoelectric
materials has been studied using the atomic pair distribution
function (PDF) method. Three candidate-models were attempted for the
structure of this class of materials using either a one-phase or a
two-phase modeling procedure. Combining modeling the PDF with HRTEM
data we show that \apst\ contains nanoscale inclusions with
composition close to \eleven\ randomly embedded in a PbTe matrix.
\end{abstract}
\pacs{61.10.-i,72.15.Jf,73.50.Lw,73.63.Bd}
\maketitle

\section{Introduction}
Compounds in the series based on composition \apst\ can exhibit
exceptional thermoelectric properties.\cite{hsu;s04} They are
promising for electrical power generation and 
in the temperature range 600 to 900 kelvin, they
are expected to significantly outperform all other reported bulk
thermoelectric systems. The dimensionless thermoelectric figure of
merit, $ZT$,\cite{ZTnote}
of the $m\sim 18$ composition material was found to reach 1.7 at 700
kelvin, compared to the highest observed ZT of only 0.84 for PbTe at
648 kelvin in n-doped
material.\cite{dugha;pb02}$^{,}$~\cite{beyer;apl02} This is a
surprisingly large enhancement in $ZT$ for the addition of just 10\%
per formula-unit of silver and antimony ions. It is clearly of the
greatest importance to trace the origin of the $ZT$ enhancement.

High resolution transmission electron microscopy (HRTEM) images from
these materials indicate the presence of nanosized domains of a
Ag-Sb-rich phase endotaxially embedded in the \seven\ matrix.\cite{quare;jacs05}
An interesting possibility is that these nanoclusters are key
components in the $ZT$~enhancement. The HRTEM images show the
clusters are randomly distributed through the matrix and are not
long-range ordered. Randomly distributed nano-scale clusters which
strain the lattice might be expected to increase phonon scattering
and reduce the thermal conductivity which would enhance $ZT$
provided the electrical conductivity was not degraded to a greater
degree. An  additional enhancement in $ZT$ is possible if the
material has an increased electronic density of states (DOS) at the
Fermi-level, $E_{f}$. A recent theoretical analysis showed that
resonant structures form in the DOS near $E_{f}$ in the presence of
ordered Ag and Sb atoms in the matrix and in the nanoclusters
observed in HRTEM.\cite{bilc;prl04}$^{,}$~\cite{humph;prl05} The
calculations used gradient corrected density functional theory and
assumed different structural models for the clusters, since details
of their structure and chemical ordering are not known. This type of
DOS resembles that of the ``best thermoelectric material" predicted
earlier.~\cite{humph;prl05,mahan;pnas96}

The composition and atomic arrangements within the nanoclusters is a
challenging topic since the clusters are not periodically long-range
ordered. They are dispersed inside a matrix and cannot be studied
crystallographically. A probing method sensitive to local structure
is needed such as the atomic pair distribution function (PDF)
analysis of x-ray powder diffraction data.\cite{egami;b;utbp03} In
the past decade the PDF technique has emerged as a powerful tool for
obtaining local structural information from complex
materials.\cite{billi;cc04,egami;b;utbp03} It is a total scattering
method that takes into account both Bragg and diffuse scattering
information and gives structural information in real space on
various length scales.  Recently, it was successfully used to study
chemical short-range ordered clusters randomly embedded in a parent
matrix,\cite{vensk;zk05} in analogy with the present situation. Here
we report a PDF study of a series of compounds in the \apst\ series
with $m=6$, 12 and 18. For comparison we also studied the end member
compound, PbTe
the $m=\infty$
member of the series. The resulting
PDFs have sufficiently high quality to see a structural signature of
the nanoclusters, even in the PbTe-rich $m=18$ compound. These differences
were sufficiently large to allow different models of the local
structure to be differentiated, confirming the existence of the
clusters in the bulk, and narrowing down their composition and the
atomic arrangement in the clusters.

\section{Experimental Details}

\subsection{Sample Preparation}
Ingots with nominal compositions \two, \four and \six\ were synthesized
by annealing, in quartz tubes under vacuum, mixtures of Ag, Pb, Sb,
and Te elements at 1000 $^\circ$C  for 8 h.  This was followed by a
fast cooling to 850 $^\circ$C  for 1 h, slow cooling to 800
$^\circ$C for 12 h, and then cooling to 400 $^\circ$C for 12 h. This
method of cooling produces more consistent samples.

\subsection{High energy x-ray diffraction experiments}
X-ray diffraction measurements were made on the \apst\ series of
materials with $m=6,$ 12, 18 and $\infty$ at room temperature using
the recently developed rapid acquisition pair distribution function
(RA-PDF) approach\cite{chupa;jac03} at the MU-CAT 6-ID-D beam-line
at the Advanced Photon Source (APS) at Argonne National Laboratory.

X-ray powder diffraction samples were prepared by carefully grinding the
compounds in a mortar and pestle and sieving through a 400-mesh
sieve.  The powders were packed into hollow flat aluminum plate
sample containers with a radius of 0.25~cm and thickness of 1.0~mm,
sealed between thin Kapton films.

The x-ray energy used was 87.005~keV ($\lambda=0.14248$~\AA). The
data were collected using a circular image plate (IP) camera Mar345,
345~mm in diameter. The camera was mounted orthogonally to the beam
path with a sample-to-detector distance of 208.86~mm which was
determined by calibrating with a silicon standard
sample.\cite{chupa;jac03}

\begin{figure}[tbp]
\centering
\includegraphics[width=0.35\textwidth,angle=0]{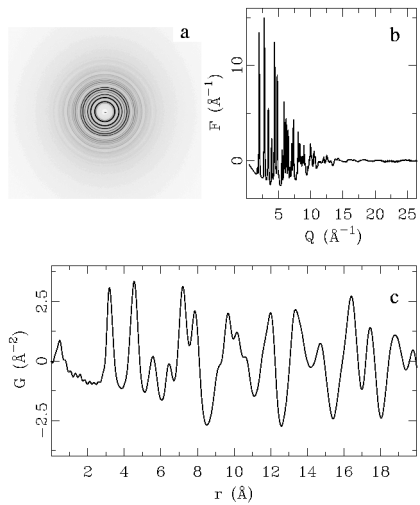}
\caption{(a) The raw data diffraction pattern observed on the image
plate.
         (b) $F(Q)$ and (c) $G(r)$ for the \six\ sample.
In the Fourier transform, $Q_{max}$ was set to 26.5~\AA$^{-1}$.}
\label{fig;lin_figure1}
\end{figure}

In order to avoid saturation of the detector, each measurement was
carried out by multiple exposure to the x-rays. Each exposure lasted
10 seconds, and each sample was exposed five times to improve the
counting statistics. An example of the raw data on the image plate
is shown in Fig.~\ref{fig;lin_figure1}(a).
All raw 2D data were integrated and converted to intensity versus
$2\theta $ format using the Fit2D program
package,\cite{hamme;esrf98} where $2\theta $ is the angle between
the incident and scattered x-rays. Data sets for the same sample
were combined using the same program. Data for the empty container
were also collected and subtracted from the sample data during the
correction step. Standard corrections for multiple scattering,
polarization, absorption, Compton scattering and Laue diffuse
scattering were applied to the integrated data to obtain the reduced
total structure function $F(Q)$, as described in detail in
Refs.~[\onlinecite{chupa;jac03,egami;b;utbp03}]. Data correction and
processing utilized the PDFgetX2 program package.\cite{qiu;jac04} An
example of the $F(Q)$ for the $m=18$ sample is shown in
Fig.~\ref{fig;lin_figure1}. Sine Fourier transformation
of $F(Q)$ gives the atomic PDF, $G(r)$, according to
$G(r)=\frac{2}{\pi}\int_{Q_{min}}^{Q_{max}} F(Q) \sin(Qr)\> dQ$,
where $Q$ is the magnitude of the scattering vector. The good
statistics in the high-$Q$ region of the data
(Fig.~\ref{fig;lin_figure1}(b)) allowed a
$Q_{max}=26.5$~\AA$^{-1}$ to be used which gives high-quality PDFs
with good resolution.  This is evident in
Fig.~\ref{fig;lin_figure1}(c) where $G(r)$ is plotted
for the sample \six.

\begin{figure}[tb]
\centering
\includegraphics[width=0.35\textwidth,angle=270]{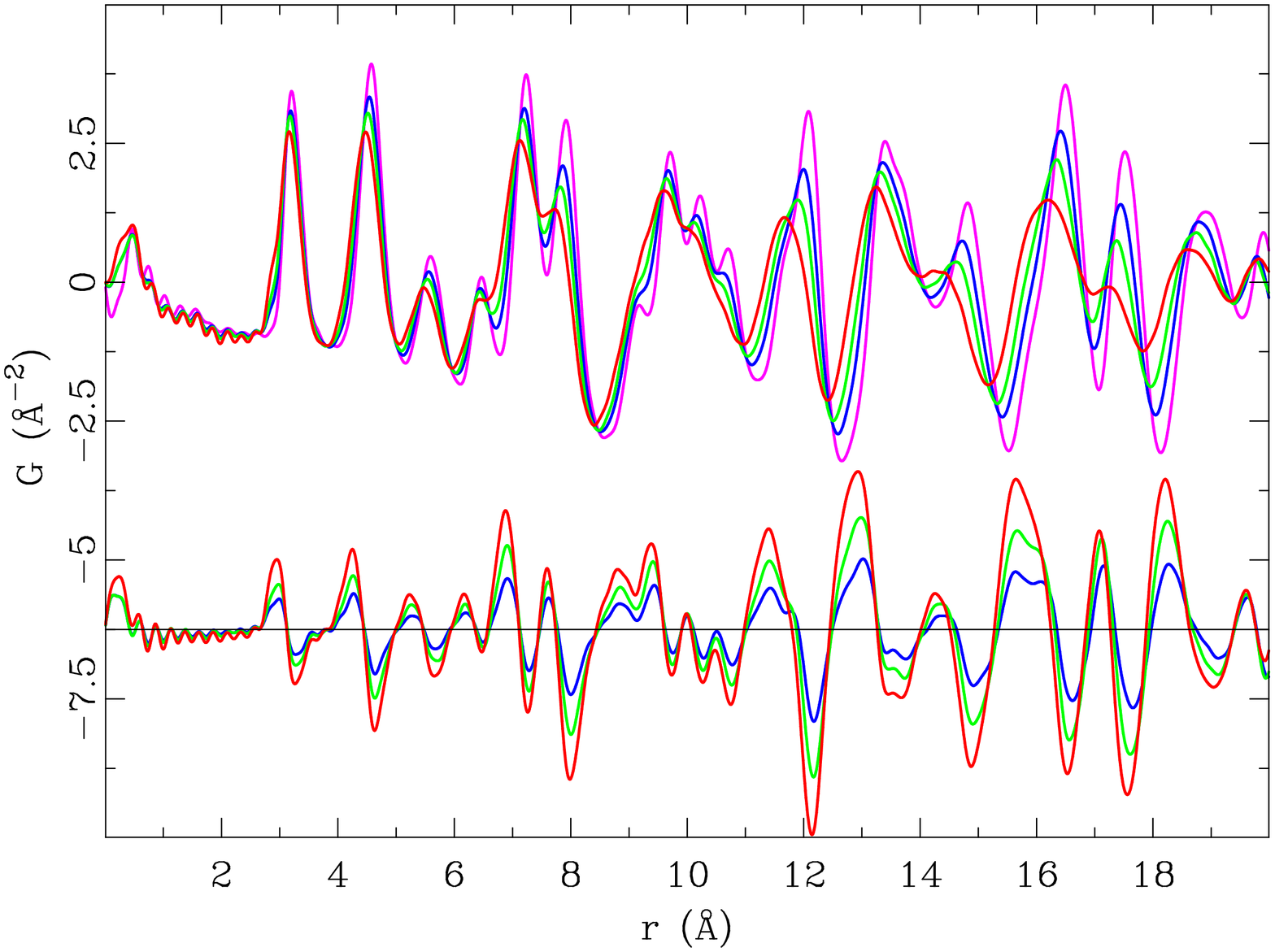}
\caption{$G(r)$ and $DG(r)$ (compared to \seven) for samples with
different $m$ value. Magenta curve is for \seven, blue curves
are for sample \six,  green for sample \four, red for \two. }
\label{fig;lin_figure2}
\end{figure}

The $G(r)$ data for all samples are plotted in
Fig.~\ref{fig;lin_figure2} on top of each other.
The difference curves plotted below are the differences between the
different $m$-value PDFs and pure PbTe. The difference curves show
fluctuations that are much larger than the estimated random errors
on the data and therefore have a real origin, encoding the local
structural differences between the \apst\ and \seven\ compounds. The
fluctuations in the difference curves are highly correlated between
the different $m$-values, growing in amplitude from $m=18$, 12 to 6,
as expected.  This suggests that the local structures in each case
are similar and gives some confidence that the results from lower
$m$-value compounds can give insight about higher $m$-members. It
also gives us confidence that the smaller ripples in the difference
curve from the $m=18$ compound have a real structural origin.

\subsection{Modeling}

Structural information was extracted from the PDFs using a
full-profile real-space local-structure refinement
method\cite{billi;b;lsfd98} analogous to Rietveld
refinement.\cite{rietv;jac69} We used an updated
version\cite{bloch;unpub05} of the program PDFFIT\cite{proff;jac99} to
fit the experimental PDFs.  PDFFIT allows for multiple data-sets to
be refined and can also handle multiple phases. Starting from a
given structure model and given a set of parameters to be refined,
PDFFIT searches for the best structure that is consistent with the
experimental PDF data. The residual function ($Rw$) is used to quantify the agreement of the
calculated PDF from model to experimental data:

\begin{equation}
Rw = \sqrt {\frac{\sum_{i=1}^{N}\omega
(r_{i})[G_{obs}(r_{i})-G_{calc}(r_{i})]^2}{\sum_{i=1}^{N}\omega
(r_{i})G_{obs}^{2}(r_{i})}}
\end{equation}
Here the weight $\omega (r_{i})$ is set to unity.

In this modeling we took advantage of the ability to refine multiple
phases in PDFFIT.  We searched for domains of Ag and Sb rich
material embedded in the PbTe matrix. Provided we fit the PDF over a
range of $r$ that is much less than the particle diameter, it is a
good approximation to model the data as being made up of two
distinct phases.  This neglects cross-terms; i.e., atom pairs where
one atom is in one phase and the neighboring atom is in the other
phase.  However, our experience suggests that these terms are small
and a reasonable and simple starting point is to neglect these terms
and model the phases as distinct (i.e., incoherent).  The HRTEM
images suggest that the nano-cluster domains have diameters of the
order of a few nanometers and our fitting is carried out over a
range up to 20~\AA . Thus, some inconsistencies in the fits in the
high-$r$ range should be attributable to the neglected cross-terms.
This approximation can be removed in the future, but only at the
expense of having to fit the data with very large models. The
success of the current modeling seems to suggest that this is not
warranted at this point.

In PDFFIT, each phase in the multi-phase mixture has its own
scale-factor that is refined. This scale factor reflects both the
relative phase-fraction of the phases, but also any differences in
the scattering power of the two phases, which depends on the
respective compositions of the phases. Here we present the equations
that allow us to extract phase fractions from the refined scale
factors of the phases. In PDFFIT, the total PDF $G(r)$ is defined as
a summation of the different phases as follows:
\begin{equation}
       G_s^\prime
       (r_{k})=f_{s}B_s(r_k)\Sigma_{p=1}^{P}{f_{p}G^s_{p}(r_{k})},
\end{equation}
where $f_{s}$ is the overall scale factor and $B_{s}$ is an
experimental resolution factor for data set $s$. The sum is over the
different structural phases, $p$, in a multiphase refinement and
$G_{p}(r_{k},s)$ is the model PDF for a single phase $p$. The
weighted abundance of each phase is given by $f_{p} =
\frac{<b_{p}>^2}{<b>^{2}}\frac{N_{p}}{N}$ where $\langle
b_{p}\rangle$ and $\langle b\rangle $ are the averaged scattering
factors for phase $p$ and the whole sample, respectively, and
$N_{p}$ and $N$ are the total atom number for phase $p$ and the whole
sample. We can easily calculate $\frac{N_{p}}{N}$ from the
stoichiometry of phase $p$ and the whole sample. After refinement we
extract $\frac{N_{p}}{N}$ from the weighted scale factor and then
compare it to the calculated one to see whether the refinement
result is self-consistent with the known stoichiometry. For example,
let's suppose we use two phases \seven\ and
\twelve\ to model \apst\ . We can set up the following two equations to
get $\frac{N_{\seven}}{N}$ and $\frac{N_{x}}{N}$:
\begin{equation}
\frac{N_{x}}{2x+4} = \frac{x N_{x}}{m(2x+4)}  + \frac{N_{\seven}}
{2m},
\end{equation}
and
\begin{equation}
N_{x} + N_{\seven} = N.
\end{equation}
Since $x$ and $m$ are known we can extract the expected ratio
$\frac{N_{p}}{N}$ for comparison with the value obtained from the
refinement.

 To test this procedure, we used a sample made by mechanically mixing
 \seven\ and \one\ powders with
atom number ratio of 1:3 and carried out a two-phase refinement. The
refined value of $\frac{N_{p}}{N}$ was 0.69 compared to the expected
values of 0.75.  This suggests that we can obtain the phase
fractions to an accuracy at the 10\% level.

Each data-set (for the finite-$m$ cases) was modeled with a sequence
of models.  Model H is a single phase homogeneous model of the
correct average composition. Models $NC0_{n}$, $NC1_{n}$, $NC2_{n}$,
$NC3_{n}$ and $NC4_{n}$ are two-phase models that test for the
presence and nature of nanoscale clusters in the material (`$NC$'
refers to nano-cluster). In all the $NC$ models, the first phase is
always a pure PbTe component. The second phase comes from the
embedded nanoclusters where we have tried different models varying
their composition and chemical ordering.  The number after the $NC$,
`0', `1', `2', `3' or `4', refers to the increasing Pb component in
the second phase as will be explained in more detail later. The
integer index $n$ refers to a different chemically ordered variant
of each nanocluster model, where $n$ increases when the chemical
ordering in the special variant increases.

\begin{figure}[tbp]
\centering
\includegraphics[width=0.5\textwidth,angle=0]{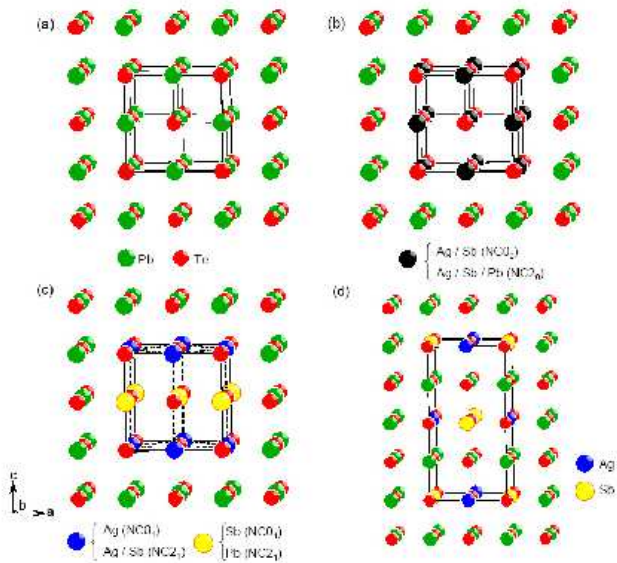}
\caption{The unit cells for different models are shown here. (a) is
the PbTe major phase. In all plots Te is shown as red  atoms and Pb
as green. (b) Chemically disordered \one\ in $NC0_{1}$ and
chemically disordered \nine\ in $NC2_{0}$. (c) Chemically ordered
\one\ in $NC0_{1}$ and partially chemically disordered \nine\ in
$NC2_{1}$. (d) Chemically ordered \nine\ in model $NC2_{2}$ resulting in
2-fold supercell along one crystal axis. In all
 models the Te (red) sublattice is not changed.}
\label{fig;lin_figure3}
\end{figure}

In solid solution model H, one homogeneous phase is defined in which the dopant
Ag, Sb atoms randomly occupy the Pb sublattice. The cubic symmetry
of the \seven\ matrix is retained, and thus only one lattice
parameter is refined. These models have four refinable
structural parameters and two experimental parameters for a total
of six refinable parameters. The PbTe structure is shown in
Fig.~\ref{fig;lin_figure3}(a).

In the case of $NC0_{n}$, a two-phase model is applied. The major phase is
still \seven . The chemical component of the second phase is the
same as bulk \one.\cite{gelle;ac59} In this model there are no Pb
atoms inside the minor phase. For the minor phase of this model we
tried both a chemically disordered cluster model $NC0_{0}$ with a
cubic unit cell and Ag, Sb atoms distributed randomly on the lead
sublattice (Fig.~\ref{fig;lin_figure3}(b)) and a tetragonal unit cell
with Ag and Sb atoms chemically ordered on the Pb sublattice sites
($NC0_{1}$, Fig.~\ref{fig;lin_figure3}(c)). These models have nine and
eleven structural parameters, respectively, resulting in eleven and
thirteen total refinable parameters.

The model $NC2_{n}$ also contains two phases, the major phase is
still \seven\ while the minor phase contains atoms with the chemical
composition of \nine. In this model, we also tried various possible
chemical ordering possibilities for the minor phases, which can be
totally chemically ordered ($NC2_{2}$), partially chemically ordered
($NC2_{1}$) and totally chemically disordered ($NC2_{0}$). In the
totally chemically ordered case, the unit cell contains 16 atoms
forming 4 layers as shown in Fig.~\ref{fig;lin_figure3}(d). There are
two types of layer. Ag, Sb and Te atoms form one type of layer and
Pb, Te atoms form the second type. The two types of layer intersect
with each other. The two lattice parameters in the plane of the
layer are the same, but the lattice parameter in the perpendicular
direction is approximately doubled. The resulting symmetry is
refined as tetragonal. This model has twelve structural parameters
and fourteen total refinable parameters.  In the $NC2_{1}$ variant,
Ag and Sb atoms distribute randomly in their plane but do not
substitute on the Pb or Te sites (Fig.~\ref{fig;lin_figure3}(c)). In the
$NC2_{0}$ case, (Fig.~\ref{fig;lin_figure3}(b)) Pb, Sb and Ag atoms
distribute randomly on the metal sublattice of the whole minor phase
resulting in a cubic structure. In both of the two latter cases,
there are only eight atoms in the unit cell. These models have
eleven and nine structural, and thirteen and eleven total refinable
parameters, respectively.

Models $NC1_{0}$, $NC3_{0}$ and $NC4_{0}$ are almost the same as
model $NC2_{0}$ except that the chemical compositions of the minor
phase are \ten, \eleven\ and \thirteen, respectively. The modeling
of the different $NC2_n$ models indicated that the PDF was not
sensitive to the degree of chemical ordering in the minor phase and
the results for chemically ordered or partially ordered cases of
models $NC1_{n}$, $NC3_{n}$ and $NC4_{n}$ are not presented here.

All refinements were performed over the range of PDF from 2.85~\AA\
to 20~\AA.  The PbTe end member compound was fit with a homogeneous
model H  and two-phase models $NC1_{n}$ and $NC2_{n}$.  All
models were fit to the $m=6$, 12 and 18 datasets.

\section{Results}
First we consider the the pure PbTe end-member compound. The
homogeneous model H, as expected, fit reasonably well resulting in
an $Rw=0.086$. Displacement parameters, $U_{iso}$, for Te and Pb
atoms are 0.013~\AA$^2$\ and 0.029~\AA$^2$\ respectively and the
lattice parameter is 6.47~\AA. Refining the two-phase model
$NC0_{n}$ and $NC2_{n}$ to the PbTe data did not result in an
improvement in $Rw$ despite the greater number of parameters. The
scale factor for the second, non-physical, phase becomes very small
(smaller than 0.3 percent), and the displacement parameters in this
phase also become very large, indicating that the fit is attempting
to eliminate the second phase. The result from the two phase
refinement shows that the PDF is able to distinguish single from
two-phase behavior.

\begin{figure}[tbp]
\centering
\includegraphics[width=0.35\textwidth,angle=0]{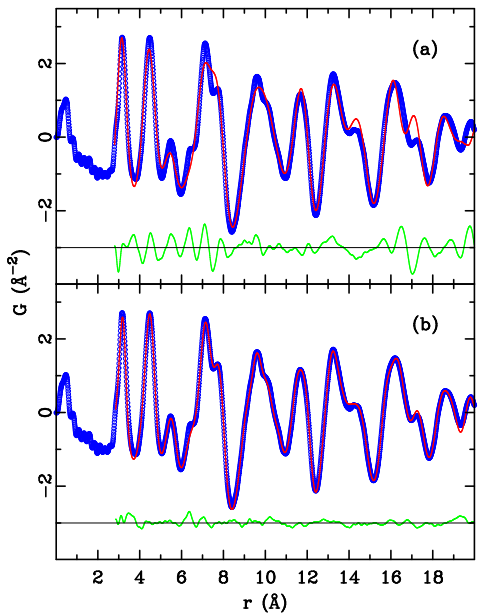}
\caption{(a) PDF from the homogeneous H model for sample \two\ . The line
with empty circles is the data, the solid line is the calculated curve
from the fitting and the line offset below is their difference. (b)
Chemically disordered case of model $NC2_{0}$ for \two . Line
attributions  are the same as in (a).}
\label{fig;lin_figure4}
\end{figure}

We now turn our attention to the $m=6$ compound that has the largest
volume fraction of second phase in it. First this was fit with the
homogeneous model H. The fit is poor as shown in
Fig.~\ref{fig;lin_figure4}(a), with $Rw=0.212$.
Significantly better fits were obtained from the two-phase models
(Table~\ref{tab;RefinementResults} and
Fig.~\ref{fig;lin_figure4}(b)) with $Rw=0.0724$ from the
chemically disordered model $NC2_{0}$.
\begin{table*}[tbp]
\centering \caption{Results from PDFFIT for the \two\ sample.
$n=\frac{N_{\seven}}{N}$ is the ratio of atom numbers in \seven\
phase to whole sample, $n_{0}$ is the expected ratio calculated from
the chemical stoichiometry (see text for details). $U_{atom}$ are
the displacement parameters for atoms on different sites.}
\begin{tabular}{|c|c|c|c|c|c|c|c|c|c|c|}
\hline  \multicolumn{2}{|c|}{} & \multicolumn{1}{c|}{model H} &
\multicolumn{1}{c|}{$NC0_{0}$} & \multicolumn{1}{c|}{ $NC0_{1}$} &
 \multicolumn{1}{c|}{ $NC1_{0}$} &
\multicolumn{1}{c|}{ $NC2_{0}$} & \multicolumn{1}{c|}{
$NC2_{1}$} & \multicolumn{1}{c|}{ $NC2_{2}$} &
\multicolumn{1}{c|}{ $NC3_{0}$} & \multicolumn{1}{c|}{ $NC4_{0}$}\\
\hline \seven  & $R_{w}$ &  0.22 &0.066 & 0.065 & 0.070 & 0.072 &
0.070 & 0.075 & 0.070 & 0.071 \\
 & $n/n_{0}$ &  -- & 0.276/0.750 & 0.257/0.75 & 0.321/0.625 &
 0.358/0.500 & 0.383/0.500 &
 0.372/0.500 &  0.376/0.325 & 0.404/0.250\\
 & $a$ &  -- & 6.41 & 6.41 & 6.41 & 6.41 & 6.41 & 6.41 &
 6.41 & 6.41 \\
 & $U_{Te}$ &  -- & 0.0291 & 0.0255 & 0.0299 & 0.0285 & 0.0305
 & 0.0297 &0.0299 & 0.0299\\
 & $U_{Pb}$ & -- & 0.0295 & 0.0307 & 0.0297 & 0.0320 & 0.0298 &
 0.0311 &  0.0294 & 0.0293 \\
\hline Phase 2 & $a$ & 6.33 & 6.22 & 6.21 & 6.22 & 6.22 & 6.23
&
6.226  & 6.219 & 6.220\\
 & $c$ &  -- & -- & 6.24 & -- & --& 6.19 & 12.40 & -- & -- \\
 & $U_{Te}$ &  0.080 & 0.0480 & 0.0488 & 0.0415 & 0.0704 & 0.0409 &
 0.0325 & 0.0384 & 0.0377 \\
 & $U_{Pb}$ &  0.061 & -- & -- & 0.08444 & 0.0550 & 0.0852 &
 0.0875 & 0.08724 & 0.0879\\
 & $U_{Ag}$ &  0.061 & 0.0782 & 0.0382 & 0.0844 & 0.0550 & 0.0852  &
 0.0875 & 0.08724 & 0.0879\\
 & $U_{Sb}$ &  0.061 & 0.0782 & 0.203 & 0.0844 & 0.0550 & 0.0852 &
 0.0875 & 0.0872 & 0.0879\\
\hline
\end{tabular}
\label{tab;RefinementResults}
\end{table*}
 The refined values are
shown in Table~\ref{tab;RefinementResults}. Similar results were
obtained from the chemically disordered models.  This analysis
strongly suggests that the Ag and Sb clusters are present in the
bulk of the material and are not an artifact of the TEM measurement.

We now wish to differentiate between the different composition
two-phase models $NC0_{n}$--$NC4_{n}$. In terms of fit to the data
and $Rw$, all four models performed comparably well, both in the
chemically ordered and disordered states. The refined parameters
that produce these good fits allow us to differentiate somewhat
between the models.  In particular, the refined phase fractions for
the two phase refinements can be compared with the values that
should be obtained based on the overall chemical composition of the
material. As can be seen in Table~\ref{tab;RefinementResults}, the
$NC0_{n}$ and $NC1_{n}$ models significantly underestimate, and $NC4_{n}$
significantly overestimates, the phase fraction.  The $NC2_{n}$ and
$NC3_{n}$ compositions give phase fractions much closer to those
expected stoichiometrically, with $NC3_{n}$ giving the best agreement.
This is strong evidence that the nanoclusters contain significant
amounts of Pb atoms and are not pure \two\ .

The refinements suggest that the average composition of the
nanoclusters is ``\eleven ". However, it is unlikely that the real
clusters have this composition since it is not possible to construct
an ordered model with this composition by interleaving Ag/Sb and Pb
layers on the Pb sublattice; it is necessary to have a layer with
Ag/Sb mixed with Pb.  As we discuss below, this is not expected on
theoretical grounds. It could come about due to the presence of
anti-phase boundaries between Pb regions and Ag/Sb regions, in
analogy with the Na$_3$BiO$_4$ material studied
previously\cite{vensk;zk05}, though it seems unlikely that this can
occur within an individual nanocluster.  From this point of view, it
seems more likely that clusters with compositions of \nine\ and
\thirteen\ coexist in the matrix yielding, on average, the observed
``\eleven " composition.  It should also be noted that some uncertainty
exists in the two-phase modeling, especially taking into account the
fact that we are modeling coherently embedded nanoclusters
approximated as an incoherent mixture.  The strong result is that
significant Pb content exists in the nanoclusters but there is
probably some uncertainty on the precise value.

We investigated the chemical ordering within the nanoclusters by
focusing on the $NC2_n$ model that lends itself to rational
chemically ordered models. Refinements of the chemically disordered
and partially ordered variants of models $NC2_n$ yielded comparable
fits to the chemically ordered fits, with comparable values of
refined parameters (Table~\ref{tab;RefinementResults}) suggesting
that the current PDF measurements alone are not sensitive enough to
differentiate the chemical ordering within the minor phase.

Finally, we note that similar results were obtained when the $m=12$
and $m=18$ samples were refined in the same way. The results for the
chemically disordered ``\eleven " model are presented in
Table~\ref{tab;RefinementResults_2}.
\begin{table*}[tbp]
\centering \caption{Results from model $NC3_{0}$ for three different
$m$-members.  $n=\frac{N_{\seven}}{N}$ is the ratio of atom numbers
in the \seven\ phase to whole sample, $n_{0}$ is the expected ratio
calculated from chemical stoichiometry. $U_{atom}$ is the thermal
factor for atoms on different site.}
\begin{tabular}{|c|c|c|c|c|c|}
\hline \multicolumn{2}{|c|}{} & \multicolumn{1}{c|}{\two} &
\multicolumn{1}{c|}{\four} & \multicolumn{1}{c|}{\six } &
\multicolumn{1}{c|}{\seven}
 \\
\hline \seven & $R_{w}$ & 0.072  & 0.066 & 0.074 & 0.086 \\
 & $n/n_{0}$ & 0.376/0.325 & 0.571/0.643 & 0.693/0.750 & - \\
 & a & 6.41 &6.43 & 6.45 & 6.47 \\
 & $U_{Te}$ &0.029 &0.0246 & 0.0214& 0.013\\
 & $U_{Pb}$ &0.032  &0.0280 &0.0262 & 0.029 \\
\hline
Phase 2 & a &6.22 & 6.26& 6.29 & - \\
& $U_{Te}$  & 0.0371 &0.0779 & 0.0826 & - \\
 & $U_{Pb}$  & 0.0815 & 0.0876 & 0.0691 & - \\
 & $U_{Ag}$  & 0.0815 & 0.0876 & 0.0691& - \\
 & $U_{Sb}$  & 0.0815 & 0.0876 & 0.0691 & - \\
\hline
\end{tabular}
\label{tab;RefinementResults_2}
\end{table*}
The refined phase fractions nicely track the nominal composition
giving us good confidence that the two-phase modeling is giving
physically meaningful results and that nanoclusters of average
composition close to \eleven\ are present.

\section{Discussion}
The success of models $NC2_{n}$ and $NC3_{n}$ verify that the TEM
observations of nanoclusters reflect a bulk average property of this
material.  These models also provide evidence for the chemical
composition of the minor phase and give a hint to the chemical
distribution of Ag, Sb and Pb atoms in the minor phase, although
little information is available about the degree of chemical
ordering.

\begin{figure}[tbp]
\centering
\includegraphics[width=0.5\textwidth,angle=0]{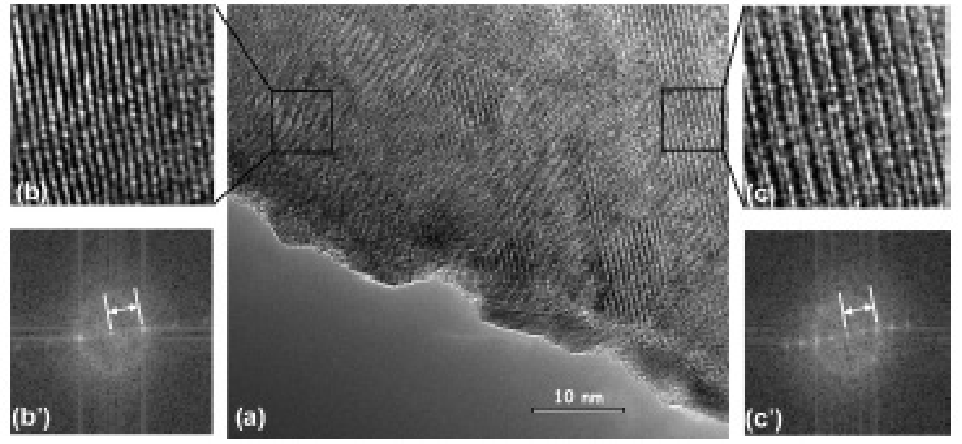}
\caption{A HRTEM image of a region of a sample of \six. The four
smaller pictures at the side are the amplified pictures for different (lattice) local region
and their fourier transformed images.}
\label{fig;lin_figure5}
\end{figure}

In Fig.~\ref{fig;lin_figure5} we show a HRTEM image that suggests that
clusters are present that result in a doubling of the lattice
parameter in the second phase, though not all clusters show this
behavior.
This is consistent with the partially or fully ordered model
$NC2_{n}$ variants, $n=1,$ 2, which alternate Pb and Ag/Sb layers on
the metallic sublattice.  The fully chemically ordered case in model
$NC2_{2}$ was found to be the stable configuration in a coulomb
lattice-gas Monte Carlo simulation study of the ground state of this
system as a function of $m$.\cite{hoang;prb05} Thus, we believe that
clusters with the totally chemically ordered form in model $NC2_{2}$
(Fig.~\ref{fig;lin_figure3}(d)) are present as nanoclusters in the
large~$m$ compounds. This may not be the unique form of the
nanoclusters, and indeed, not all the nanoclusters evident in the
TEM images show this cell doubling. They presumably form by a
nano-phase separation of constituents accompanied by an imperfect
and defective ordering and there appears to be considerable spatial
disorder of the chemical constituents; though the nanoclusters are
coherently endotaxially embedded in the matrix, they are not well ordered.
Incorporating more Pb in the nanoclusters allows the system to
balance its desire to phase separate, with maintaining a degree of
lattice matching to keep the nanoparticles embedded in the matrix
without incoherent interfaces.  The refined lattice parameters for
the nanocluster phases are smaller than the matrix: 6.23-6.29~\AA\
compared to 6.41~\AA\ for the strained matrix and 6.47~\AA\ for
relaxed PbTe.  Both the chemical inhomogeneities and the
inhomogeneous lattice strain are likely to increase phonon
scattering.  The size of the nanoclusters may also be important in
making this scattering mechanism effective.  The short-range nature
of the local chemical ordering will broaden out Fermi-surface
resonances shown to be important in thermopower
enhancement\cite{bilc;prl04}. Presumably, the size of the nanoclusters,
their exact composition,the atomic ordering within them and their
concentration with \seven\ will be a sensitive function of the
preparation conditions.

Adding Pb atoms in the minor phase greatly improves the result of
the refinement. The reason is that Pb atomic number is much larger
than the atomic numbers of Ag, Sb and Te and its scattering factor is
quite different from those of the other three.

\section{summary}
In this structural study based on the PDF method, we verified that
in the bulk material of \apst, nanoclusters of a minor phase
containing Ag, Pb, Sb and Te atoms form in the matrix of \seven. We
give evidence showing that the chemical composition of the minor
phase is most likely between \nine\ and \thirteen.  We propose a structure for the minor
phase based on PDF, TEM and theoretical considerations.

\begin{acknowledgements} We would like to acknowledge help from Didier
Wermeille, Doug Robinson, Pavol Juhas, Gianluca Paglia, Ahmad
Masadeh, Hasan Yavas, HyunJeong Kim and Mouath G. Shatnawi in
collecting data. We also thank S.D.Mahanti, Khang Hoang for helpful
discussions. Work in the Billinge group was supported by the NSF
NIRT grant DMR-0304391. Work in the Kanatzidis group was supported by the
ONR(MURI) program. Data were collected at the 6IDD beam-line
Advanced Photon Source (APS). Use of the APS is supported by the
U.S. DOE under Contract No. W-31-109-Eng-38. The MUCAT sector at the
APS is supported by the U.S. DOE under Contract No. W-7405-Eng-82.
\end{acknowledgements}

\begin{thebibliography}{10}
\bibitem{ZTnote}
$ZT= \sigma S^{2} T /\kappa $, where $T$ is the operating temperature of the
  device, $S$ is the thermopower, and $\sigma$ and $\kappa$ are the electrical
  and thermal conductivities of the material, respectively.
\end{thebibliography}

\end{document}